\documentstyle[12pt,fleqn]{article}
\begin{document}
\begin{flushright}
{MANIT-11-94}\\
{gr-qc/9507026}
\end{flushright}
\begin{center}
{\Large Heat Kernel and Loop Currents by the Generating-Function Method
}
\footnote{Talk at ``Heat kernel techniques and
quantum gravity'' conference, Winnipeg, Canada, August 1994.}
\\
\vspace{3mm}
{\sl Andrei Barvinsky}
\footnote{Present address: Lebedev Physics Institute,
Leninsky pr. 53, Moscow  117924, Russia;
e-mail: andrei@rq1.fian.msk.su}
\\
\vspace{5mm}
Department of Physics\\
University of Alberta\\
Edmonton, Canada \\
\vspace{5mm}
and \\
\vspace{10mm}
{\sl Yuri Gusev}
\footnote{E-mail: gusev@cc.umanitoba.ca}
\\
\vspace{3mm}
Department of Physics\\
University of Manitoba\\
Winnipeg, Canada
\\
\vspace{5mm}
{\bf Abstract}\\
\end{center}
{\small
The generating function method is  applied to the
trace of the heat kernel and the one-loop
effective action derived from the covariant perturbation
theory.
The basis of curvature invariants of second order
for the heat kernel (Green function) is built and
simple rules for form factor manipulations
are proposed. The results are checked by deriving the
Schwinger-DeWitt series of the heat kernel
and divergences of
one-loop currents.
}

\section{The generating function method}

The present talk reviews a progress on calculations of
heat kernel and relevant quantum objects in the
framework of the covariant perturbation theory \cite{cpt,cpt4,JMP94}.
As generating functions we take
the one-loop effective action
\begin{equation}
W=-\frac{1}{2}\int^{\infty}_{0}\frac{ds}{s} \, {\rm Tr} K(s)  \label{efac}
\end{equation}
and the functional trace of the heat kernel
\begin{equation}
{\rm Tr} K(s)=
\int  \! dx \, g^{1/2}(x) \, {\rm tr} \hat{K}(s|x,x) \label{traceK}
\end{equation}
while the heat kernel itself
\begin{equation}
\hat{K}(s|x,y)={\rm e}^{ s \hat{F}(\nabla)}\delta(x,y)  \label{kernel}
\end{equation}
is defined by the second order operator
\begin{equation}
F=\Box\hat{1}+\hat{P}-\frac{1}{6}R\hat{1},
\ \ \
\Box=g^{\mu\nu}\nabla_{\mu}\nabla_{\nu}                     \label{oper}
\end{equation}
acting on any column of fields $\varphi^{a}(x)$, an overhat indicating the
matrix-valued nature of the corresponding field objects:
$\hat{1}=\delta^{a}_{b}, \hat{P}=P^{a}_{b}$ etc. This operator contains
 three independent fields: the spacetime metric $g^{\mu\nu}$, the
 arbitrary connection entering the covariant derivative $\nabla_{\mu}$ and
 the potential term $\hat{P}$ which correspond to the three curvatures:
Riemann tensor
\footnote{
               Our conventions are:
               $R=g^{\mu\nu} R_{\mu\nu}=
                g^{\mu\nu}R^{\lambda}_{.\mu\lambda\nu}=
               g^{\mu\nu}(\partial_{\lambda}
                \Gamma^{\lambda}_{\mu\nu}-...)$.
                }
$R_{\mu\nu\alpha\beta}$,
the potential term $\hat{P}$, and
the commutator curvature
\begin{equation}
                   (\nabla_{\mu}\nabla_{\nu}-
                  \nabla_{\nu}\nabla_{\mu})\varphi^A=  \label{commutator}
                     {\cal R}^A_{B\mu\nu}\varphi^B.
\end{equation}

Our purpose is to derive  coincidence limits of the heat kernel
of the most general form
\begin{equation}
   \nabla^x \ldots \nabla^x \nabla^y \ldots \nabla^y
   \hat{K}(s|x,y)|_{y=x}               \label{D..DK}
\end{equation}
that bear all essential information of quantum field theory
\cite{GAVStrasburg,DeWittBook,BV85}.
     From (\ref{D..DK}) Green's functions are defined via
Schwinger's
 proper-time representation \cite{DeWittBook,BV85}
\begin{equation}
 G(x,y)\equiv
 \nabla^x\! \ldots \nabla^x \nabla^y\! \ldots \nabla^y
\frac{\hat{1}}{F(\nabla)}\delta(x,y)=-\!\int^{\infty}_0 \!ds \label{Greenf}
 \nabla^x \! \ldots \nabla^x \nabla^y \! \ldots \nabla^y
\hat{K}(s|x,y).
\end{equation}

Actually our final  goal is a computation
of two-loop effective action \cite{BV87,VdV92}
\vspace{-6mm}
\begin{equation}
W_{\rm two-loop}=
-\frac12 \;
\hbox{
\begin{picture}(40,40)
\put(20,0){\circle{40 } }
\put(0,0){\circle*{3} }
\put(40,0){\circle*{3} }
\put(0,0){\line(1,0){40} }
\end{picture}
}
\;-\frac18 \:
\hbox{
\begin{picture}(80,40)
\put(20,0){\circle{40 } }
\put(40,0){\circle*{3} }
\put(60,0){\circle{40} }
\end{picture}
}
\label{two-loop} \end{equation}
\vspace{6mm}
for Yang-Mills and gravity theories.
Building blocks  for ``fish'' and ``eight''
diagrams (\ref{two-loop})
are  ``tadpoles'' (universal functional traces \cite{BV85,BV87})
\vspace{-6mm}
 \begin{equation}
G(x,x)=
\hspace{5mm}
\hbox{
\begin{picture}(60,40)
\put(25,0){\circle{40 } }
\put(45,0){\circle*{3} }
\put(45,-2){ \it x}
\end{picture}
}
                                   \label{UFT}
\end{equation}
\vspace{7mm}
and polarization operators
\vspace{-6mm}
\begin{equation}
G_1(x,y) G_2(x,y)=\hspace{5mm}
\hbox{
\begin{picture}(60,40)
\put(25,0){\circle{40 } }
\put(5,0){\circle*{3} }
\put(-7,-2){ \it x}
\put(45,0){\circle*{3} }
\put(45,-2){ \it y}
\end{picture}
}
                                           \label{PO}
\end{equation}
\vspace{7mm}
Having these two sets of basic elements the two-loop
graphs can be easily composed,
e.g.
\vspace{-3mm}
\[
\int dx \, G_1(x,x)\, G_2(x,x) =
\hbox{
\begin{picture}(80,40)
\put(20,0){\circle{40 } }
\put(40,0){\circle*{3} }
\put(60,0){\circle{40} }
\end{picture}
}
\]
\vspace{7mm}

Briefly the generating expressions  approach
originates from a simple principle \cite{BGusev92}
\[
\frac1s \delta \Big( {\rm Tr} \, {\rm e}^{s\hat{F}}\Big)
={\rm Tr} \Big(\delta \hat{F} {\rm e}^{s\hat{F}}\Big)
\]
which implies
\begin{equation}
\frac{1}{s}\delta \Big({\rm Tr} K(s)\Big)
= \int dx \, {\rm tr}\Big[
\delta \hat{F}(\nabla^x)
\hat{K}(s|x,y)\Big]\Big|_{y=x}.    \label{var}
\end{equation}
Because there are three independent curvatures in the
differential operator (\ref{oper})
two variational equations can be written down immediately:
\begin{eqnarray}
&&
\hat{K}(s|x,x)=
\frac{1}{s}\frac{\delta}{\delta \hat{P}(x)}
{\rm Tr} {K}(s),
                               \label{varP}
\\
&&
\nabla_{\alpha}      \hat{K}(s|x,y)|_{y=x}   =
\frac{1}{s}  \frac{\delta}{\delta \hat{\Gamma}^{\alpha}(x)}
{\rm Tr}  \, {\rm e}^{s(F+\Gamma^{\mu}{\nabla}_{\mu})}
\Big|_{\Gamma^{\mu}=0},
                                     \label{varGamma}
\end{eqnarray}
where a new operator $\hat{F} \rightarrow \hat{F} +
                            \hat{\Gamma}^{\mu} \nabla_{\mu}$
in (\ref{varGamma}) is generated by redefinition of
the covariant derivative connection
$\nabla_\mu \rightarrow \nabla_\mu +\hat{\Gamma}_\mu$,
consequently  $\hat{P}$ and $\hat{\cal R}_{\mu\nu}$
are redefined.
Third variational equation over the metric $g^{\mu\nu}$
give us a combination of
$ \nabla_{(\alpha} \nabla_{\beta)} {\rm tr} K(s)$
and $ g_{\alpha\beta} \Box  {\rm tr} K(s)$,
so the desired object
\begin{equation}
 \nabla_{\alpha} \nabla_{\beta}\hat{K}(s|x,y)|_{y=x}  \label{DDK}
\end{equation}
can be expressed using the outcome of (\ref{varP})
and the heat equation itself
\[
\frac{\partial}{\partial s}\hat{K}(s)=
\hat{F}(\nabla)\hat{K}(s).
\]
An antisymmetric part
                   $\nabla_{[\alpha} \nabla_{\beta]}
                   \hat{K}(s)$
is recovered from
(\ref{commutator}) and  (\ref{varP}).
Fortunately, the terms in (\ref{DDK})
removed by the matrix trace operation are  absent in first order
as proved by the covariant perturbation theory \cite{BGusev92}.

The method of generating expressions
first has been proposed  \cite{BV87}
for analysis of divergences of $G(x,x)$
and $G_1(x,y) G_2(x,y)$.
However, even local expansions of two-loop
diagrams require knowledge of finite
nonlocal structures of one-loop generating
functions  \cite{BV87}; thus we need to resort
to the covariant perturbation theory.

The covariant perturbation theory allows one to
 obtain an  expansion of (the trace of) the  heat kernel in powers of the
curvatures
to a given accuracy
             ${\rm O}[\Re^n]$
with a  collective notation
             $ \Re=(R^{\alpha\beta\mu\nu},
                          \hat{\cal R}_{\mu\nu},
                          \hat{P})$.
In Refs. \cite{cpt,cpt4,JMP94} ${\rm Tr} K(s)$ and $W$ were calculated in a
 $2\omega$-dimensional asymptotically flat Euclidean spacetime to third
 order in curvature
 \begin{eqnarray}
{\rm Tr} K(s) &=& \frac{1}{(4 \pi s)^\omega}\int\! dx\,
g^{1/2}\,{\rm tr}\Big\{\hat{1}+s\hat{P}
+s^2\sum^{5}_{i=1}f_{i}(-s\Box_2)\Re_1\Re_2(i)
\nonumber\\&&\mbox{}
+s^3\sum^{11}_{i=1}F_{i}(-s\Box_1,-s\Box_2,-s\Box_3) \Re_1 \Re_2 \Re_3 (i)
\nonumber\\&&\mbox{}
+s^4\sum^{25}_{i=12}F_{i}(-s\Box_1,-s\Box_2,-s\Box_3) \Re_1 \Re_2 \Re_3 (i)
\nonumber\\&&\mbox{}
+s^5\sum^{28}_{i=26}F_{i}(-s\Box_1,-s\Box_2,-s\Box_3) \Re_1 \Re_2 \Re_3 (i)
\nonumber\\&&\mbox{}
+s^6 F_{29}(-s\Box_1,-s\Box_2,-s\Box_3) \Re_1 \Re_2\Re_3(29)
+{\rm O}[\Re^4]\big\},   \label{TrK3}
\end{eqnarray}
where quadratic tensor structures
and terms of third order in curvature
make up a basis
of nonlocal curvature invariants
\cite{cpt4,JMP94,GAVStrasburg},
and the notation on the left-hand side of (\ref{TrK3})
assumes that $\Box_i$ acts on $\Re_i$.

The only important feature of the form factors
$f_i$ and $F_i$ required here is that they are functions
of the operator
\[
\xi_i=-s\Box_i
\]
and do not depend explicitly on
the curvatures $\Re$.
This is an artifact of the method
which always destroys
one curvature reducing an accuracy ${\rm O}[\Re^n]$
by one order.
In this circumstance
the variation of form factors is not required.

The covariant perturbation theory represents the high-energy
approximation so its validity is restricted by the condition
\[  \nabla \nabla {\Re} >> \Re^2.\]

All results of next two sections
obtained with the generating function methods have been
checked by the covariant perturbation theory formulae \cite{cpt}.


 \section{The heat kernel to second order in curvatures}

We shall not elaborate on  results for $\nabla_{\alpha} \hat{K}(s)$
and $\nabla_{\alpha}\nabla_{\beta} \hat{K}(s)$
\cite{BGusev92},
but will consider  the heat kernel $\hat{K}(s)$ in more detail.
One can easily observe that only twelve of 29 cubic tensor structures
(\ref{TrK3}) contain the matrix $\hat{P}$ and therefore contribute
to the  heat kernel.
The variational derivative of
${\rm Tr} K(s)$ over $\hat{P}$ results in
the following expression
\begin{eqnarray}
\hat{K}(s)&=&
              \frac{1}{(4\pi s)^\omega}\, g^{1/2}
                        \Big\{ \hat{1}+
                   s\Big(g_1(-s\Box)\hat{P}
+g_2 (-s\Box)R\hat{1}\Big)\nonumber\\&&
\mbox{}+s^2 \sum_{i=1}^5 G_i(-s\Box_1,-s\Box_2,-s\Box_3)
                                \Re_1 \Re_2 [i]
\nonumber\\ \mbox{}&&
+  s^3  \sum_{i=6}^{10}
                   G_i(-s\Box_1,-s\Box_2,-s\Box_3)
                                 \Re_1 \Re_2 [i]
\nonumber\\ \mbox{}&&
+ s^4  G_{11}
              (-s\Box_1,-s\Box_2,-s\Box_3)
                      \Re_1 \Re_2 [11]     \label{K2}
+{\rm O}[\Re^3]\Big\},
\end{eqnarray}
where first order form factors are formed by the second order
ones of (\ref{TrK3}) as
\begin{eqnarray}
&&
g_1(\xi)=2 f_4(\xi),
\nonumber\\&&
g_2(\xi)= f_3(\xi),
\end{eqnarray}
and quadratic tensor structures $\Re_1 \Re_2 (i)$  are
\begin{eqnarray}
&&\Re_1 \Re_2 [1]= \hat{P}_1 \hat{P}_2,
\nonumber\\&&
\Re_1 \Re_2 [2]=  \hat{\cal R}^{\mu\nu}_1 \hat{\cal R}_{2 \mu \nu},
\nonumber\\&&
\Re_1 \Re_2 [3]= \hat{P}_1 R_2,
\nonumber\\&&
\Re_1 \Re_2 [4]= R_1 R_2 \hat{1},
\nonumber\\&&
\Re_1 \Re_2 [5]= R^{\mu\nu}_1 R_{2\mu\nu}\hat{1},
\nonumber\\&&
\Re_1 \Re_2 [6]=
\nabla_{\mu}\hat{\cal R}^{\mu\nu}_1
\nabla^{\alpha}\hat{\cal R}_{2\alpha\nu},
\nonumber\\&&
\Re_1 \Re_2 [7]=
\big[\nabla_{\alpha}\hat{P}_1,
\nabla_{\beta}\hat{\cal R}_2^{\beta\alpha}\big],
\nonumber\\&&
\Re_1 \Re_2[8]= \nabla_\mu \nabla_\nu \hat{P}_1 R_2^{\mu\nu},
\nonumber\\&&
\Re_1 \Re_2 [9]=\nabla_\alpha R_{1\mu\nu}
\nabla^{\mu}R_2^{\nu\alpha}\hat{1},
\nonumber\\&&
\Re_1 \Re_2 [10]=\nabla_\mu \nabla_\nu R_1 R_2^{\mu\nu}\hat{1},
\nonumber\\&&
\Re_1 \Re_2 [11]= \nabla_\alpha \nabla_\beta R_{1\mu\nu}
\nabla^\mu \nabla^\nu R_2^{\alpha\beta}\hat{1}.           \label{RRK2}
\end{eqnarray}
Without gravity, the basis (\ref{RRK2})
reduces to only four curvature structures.
Taking into account the symmetries \cite{cpt4}
of form factors
(\ref{TrK3}) the second order form factors $G_i, i=1$ to 11
are expressed
via $F_i$
in the following way,
\begin{eqnarray}
&&G_1({\xi_1},{\xi_2},{\xi_3})=
3F_1({\xi_1},{\xi_2},{\xi_3}),
\nonumber\\&&
G_2({\xi_1},{\xi_2},{\xi_3})=
F_3({\xi_1},{\xi_2},{\xi_3}),
\nonumber\\&&
G_3({\xi_1},{\xi_2},{\xi_3})=
F_6({\xi_1},{\xi_2},{\xi_3})
|_{{\xi_2} \leftrightarrow {\xi_3}}
+F_6({\xi_1},{\xi_2},{\xi_3})
|_{{\xi_1} \rightarrow {\xi_3},{\xi_3} \rightarrow {\xi_2},{\xi_2}
\rightarrow {\xi_1}}
\nonumber\\&&\ \ \ \ \mbox{}
-\Big(\frac12 {\xi_2}+\frac12 ({\xi_3}-{\xi_2}-{\xi_1})\Big)
F_{17}({\xi_1},{\xi_2},{\xi_3})
|_{{\xi_1} \rightarrow {\xi_2},{\xi_2} \rightarrow {\xi_3},{\xi_3}
\rightarrow {\xi_1}},
\nonumber\\&&
G_4({\xi_1},{\xi_2},{\xi_3})=
F_4({\xi_1},{\xi_2},{\xi_3})+\frac14 ({\xi_3}-{\xi_2}-{\xi_1})
F_{15}({\xi_1},{\xi_2},{\xi_3}),
\nonumber\\&&
G_5({\xi_1},{\xi_2},{\xi_3})=
F_5({\xi_1},{\xi_2},{\xi_3}),
\nonumber\\&&
G_6({\xi_1},{\xi_2},{\xi_3})=
F_{14}({\xi_1},{\xi_2},{\xi_3}),
\nonumber\\&&
G_7({\xi_1},{\xi_2},{\xi_3})=
-F_{13}({\xi_1},{\xi_2},{\xi_3})|_{{\xi_1} \leftrightarrow {\xi_2}},
\nonumber\\&&
G_8({\xi_1},{\xi_2},{\xi_3})=
F_{17}({\xi_1},{\xi_2},{\xi_3})|_{{\xi_1} \leftrightarrow {\xi_2}}+
F_{17}({\xi_1},{\xi_2},{\xi_3})
|_{{\xi_1} \rightarrow {\xi_2},{\xi_2}
\rightarrow {\xi_3},{\xi_3} \rightarrow {\xi_1}},
\nonumber\\&&
G_9({\xi_1},{\xi_2},{\xi_3})=
F_{16}({\xi_1},{\xi_2},{\xi_3}),
\nonumber\\&&
G_{10}({\xi_1},{\xi_2},{\xi_3})=
-F_{15}({\xi_1},{\xi_2},{\xi_3})|_{{\xi_1} \leftrightarrow {\xi_2}},
\nonumber\\&&
G_{11}({\xi_1},{\xi_2},{\xi_3})=
F_{26}({\xi_1},{\xi_2},{\xi_3}).
\end{eqnarray}

It should be emphasized again that these rules are applicable to
any of the tabulated form factors of Ref. \cite{cpt4} related to
the heat kernel trace; i.e.,
they can be in the explicit or $\alpha$-polynomial representations,
or even large and short time expansions.
Due to the special status  of the matrix $\hat{P}$,
which is a curvature and perturbation  simultaneously,
the variation (\ref{varP}) was performed at the level of
perturbations of the metric and covariant derivative as well.

The explicit form of a few first form factors is as follows
\begin{eqnarray}
g_1&=&f(\xi),
\\
g_2&=&\frac{1}{12}f(\xi)+\frac12\frac{f(\xi)-1}{\xi},
\\
G_1&=& F({\xi_1},{\xi_2},{\xi_3}),   \label{G1}
\\
G_2&=& F({\xi_1},{\xi_2},{\xi_3})
\Big[ \frac{2 {\xi_1} {\xi_2}}{\Delta^2}
({\xi_3}+{\xi_2}-{\xi_1}) ({\xi_3}+{\xi_1}-{\xi_2})
+ \frac{2}{\Delta}({\xi_3}-{\xi_2}-{\xi_1})\Big]
\nonumber\\ &&
\mbox{}
 - f({\xi_1})\frac{4 {\xi_1} {\xi_2}}{\Delta^2}
({\xi_3}+{\xi_1}-{\xi_2})
 - f({\xi_2})\frac{4 {\xi_1} {\xi_2}}{\Delta^2}
({\xi_3}+{\xi_2}-{\xi_1})
\nonumber\\ &&
\mbox{}
 -f({\xi_3})\frac{1}{\Delta^2}
(-6 {\xi_1} {\xi_2} {\xi_3}
-3 {\xi_1} {\xi_3}^2
-3 {\xi_2} {\xi_3}^2
+3 {\xi_3} {\xi_1}^2
+3 {\xi_3} {\xi_2}^2
\nonumber\\&&
 \mbox{}
+{\xi_3}^3
+{\xi_1} {\xi_2}^2
+{\xi_2} {\xi_1}^2
-{\xi_2}^3
-{\xi_1}^3),                  \label{G2}
\end{eqnarray}
where
$f(\xi)$ is the basic second-order form factor
\cite{cpt,cpt4}
\begin{eqnarray}
    f(\xi)  &\equiv&
     \Big<{\rm exp} (- \alpha_1 \alpha_2 \xi ) \Big>_2 \equiv
    \nonumber\\ &&\mbox{}
     \int_{ \alpha   \geq 0 }    \! d^2  \alpha \,
     \delta(1-\alpha_1-\alpha_2)\exp(-\alpha_1\alpha_2\xi)=
     \int^1_0\!d\alpha\,{\rm e}^{-\alpha(1-\alpha)\xi} \label{smallf}
\end{eqnarray}
and
$ F_{i}(\xi_1,\xi_2,\xi_3)$
 --  the basic third-order form factor \cite{cpt4,JMP94}
\begin{eqnarray}
&& F({\xi_1},{\xi_2},{\xi_3})  \equiv
\Big<    {\rm e }^{  \Omega  }  \Big>_3  \equiv
\int_{\alpha   \geq 0}    d^3   \alpha \,
 \delta(1-\alpha_1-\alpha_2-\alpha_3)
\exp(\Omega),
\\ &&\mbox{}
\Omega =
-\alpha_1\alpha_2{\xi_3}
-\alpha_2\alpha_3{\xi_1}
-\alpha_1\alpha_3{\xi_2}
\end{eqnarray}
with the  polynomial
\begin{equation}
\Delta={{\xi_1}}^2+{{\xi_2}}^2+{{\xi_3}}^2
-2{\xi_1}{\xi_2}-2{\xi_1}{\xi_3}-2{\xi_2}{\xi_3}.         \label{Delta}
\end{equation}
The
statement of the full set of form factors
takes several pages
and will be published elsewhere.
A noncovariant perturbation theory first was formulated
for a scalar field  in Ref. \cite{WFO81}
where $g_1$ and $G_1$ can be found.

The apparent way to verify (\ref{K2})
is to implement the functional trace operation (\ref{traceK}).
Unfortunately we can not manage this in the  given
form factor representation  (\ref{G1})--(\ref{G2})
because of  the presence of $\Delta$  in
denominators.  Only representation
with help of $\alpha$-polynomials
which naturally arises from the covariant perturbation theory
and, in fact, serves as a starting point for the explicit
representation
\cite{cpt4} can be used for this purpose.
In this representation form factors admit significantly more
compact form
\begin{eqnarray}
g_1&=&
                \left<
                         {\rm e}^{-\alpha_1\alpha_2 \xi }
                \right>_{2},   \label{g1a}
\\
g_2&=&
               \left<
                    \left(
                           \alpha_1^2-\frac16
                    \right){\rm e}^{-\alpha_1\alpha_2 \xi }
                 +\frac{{\rm e}^{-\alpha_1\alpha_2 \xi }-1}{\xi}
               \right>_{2},   \label{g2a}
\\
G_1&=&\left< \rm{e}^{\Omega}\right>_3, \label{G1a}
\\
G_2&=&\left< 2{\alpha_1}{\alpha_2} \label{G2a}
\rm{e}^{\Omega}\right>_3.
\end{eqnarray}
The $\alpha$-representation (\ref{g1a}-\ref{G2a})
is not unique, i.e. some form factors $G_i$, that  in fact vanish,
can present in the heat kernel \cite{cpt4}.
Due to this fact
there is an
additional
quadratic structure linear in ${\cal R}_{\mu\nu}$
\[
\Re_1\Re_2[12]=\nabla_{\nu}{\cal R}_1^{\nu\mu}\nabla_{\mu}R_2
\]
absent  in the final expression (\ref{K2}).
All integrals over the  whole spacetime
should be discarded and all second order form factors reduced to
first order ones by identities like
\begin{equation}
{\rm tr}\!
\int d x \,  g^{1/2}(x) F(\xi_1, \xi_2, \xi_3)
\Re_1 \Re_2=
\frac12 {\rm tr}\!
\int  d x  \, g^{1/2}(x)
\Re f(\xi) \Re +{\rm O}[\Re^3].
\end{equation}
We have verified that the form factor tables of (\ref{K2})
are in a full agreement with ones of ${\rm Tr} K(s)$ (\ref{TrK3}).

      From the variational principle and (\ref{traceK})
we can conclude that there exists a remarkable link between
two neighboring
 orders in the curvature of the heat kernel trace, namely,
each lower order is completely defined by the higher order
\begin{equation}
                           {\cal K}_{n-1}=
                   {\rm tr} \! \int  dx   \,
   \frac{\delta}{\delta \hat{P}} \, {\cal K}_n, \label{K-K}
\end{equation}
where
\[
{\cal K}_n \equiv
{\rm tr} \! \int  dx \, g^{1/2}
\sum_i F_i(\nabla_1, \ldots , \nabla_n)
\Re_1 \ldots  \Re_n(i).
\]

Another good check is short time expansions
for the form factors of the heat kernel.
They can be treated in either explicit form or the form
of $\alpha$-representation. Of course, the expanded
form factors are nonlocal but the
local Schwinger-DeWitt coefficients can be restored
using the identity \cite{cpt4,JMP94}
\begin{eqnarray}
         \Box R^{\alpha\beta\mu\nu}
                                                      &\equiv&
      \frac12\Big(
\nabla^\mu \nabla^\alpha R^{\nu\beta}
+\nabla^\alpha \nabla^\mu R^{\nu\beta}
-\nabla^\nu \nabla^\alpha R^{\mu\beta}
-\nabla^\alpha \nabla^\nu R^{\mu\beta}
                \nonumber\\&&\mbox{}
-\nabla^\mu \nabla^\beta R^{\nu\alpha}
-\nabla^\beta \nabla^\mu R^{\nu\alpha}
+\nabla^\nu \nabla^\beta R^{\mu\alpha}
+\nabla^\beta \nabla^\nu R^{\mu\alpha}
       \Big)
                 \nonumber\\&&\mbox{}
+ R^{[\mu}_{\lambda} R^{\nu] \lambda\beta\alpha}
+ R^{[\alpha}_{\lambda} R^{\beta] \lambda\nu\mu}
- 4  R^{\alpha\ [\mu}_{\  \sigma \ \ \lambda}  R^{\nu] \lambda\beta\sigma}
- R^{\alpha\beta}_{\ \ \ \sigma\lambda} R^{\mu\nu\sigma\lambda}.
\end{eqnarray}

What we get finally takes the shape of
the local Schwinger-DeWitt series \cite{DeWittBook,BV85}
\begin{equation}
\hat{K} (s) = \frac{g^{1/2}}{(4 \pi s)^{\omega}}
                     \sum_{n=0}^{\infty} s^n \hat{a}_n (x,x)   \label{SDs}.
\end{equation}
Terms $\hat{a}_n$ up to $n=3$  are well-established
\cite{DeWittBook,BV85,Gilkey75,Avramidi}
but the fourth coefficient is of special interest.
Now available results for $\hat{a}_4$ are to be futher reduced
\cite{Avramidi} or restricted to scalar field case
\cite{WFO81,AmsterBO89}
lacking gauge fields and matrix structure itself.
Therefore even limited form  $({\rm O}[\Re^3])$ of $\hat{a}_4$
deserves to be reproduced completely here
\begin{eqnarray}
&&      \hat{a}_4     (x,x)    =
\frac{1}{840}{\Box}^3\hat{P}+\frac{1}{15120}{\Box}^3 R\hat{1}
                \nonumber\\&&\mbox{}
+\frac{1}{360} {\Box}( \hat{P}\Box \hat{P})
+\frac{1}{360}\Box \hat{P}\Box \hat{P}
+\frac{1}{360}\Box(\Box\hat{P} \hat{P})
                 \nonumber\\&&\mbox{}
+\frac{1}{360}{\Box}^2(\hat{P} \hat{P})
+\frac{1}{360}{\Box}^2\hat{P} \hat{P}
+\frac{1}{360} \hat{P}{\Box}^2 \hat{P}
               \nonumber\\&&\mbox{}
+\frac{1}{3360}{\Box}^2
\hat{\cal R}^{\mu\nu} \hat{\cal R}_{\mu \nu}
+\frac{1}{3360}
\hat{\cal R}^{\mu\nu}{\Box}^2 \hat{\cal R}_{\mu \nu}
+\frac{1}{2520}
{\Box}\hat{\cal R}^{\mu\nu}{\Box} \hat{\cal R}_{\mu \nu}
                \nonumber\\&&\mbox{}
+\frac{1}{1680}
{\Box}({\Box}\hat{\cal R}^{\mu\nu} \hat{\cal R}_{\mu \nu})
+\frac{1}{1120}
{\Box}^2(\hat{\cal R}^{\mu\nu} \hat{\cal R}_{\mu \nu})
+\frac{1 }{1680}
{\Box}(\hat{\cal R}^{\mu\nu} {\Box}\hat{\cal R}_{\mu \nu})
                  \nonumber\\&&\mbox{}
+\frac{1}{15120}{\Box}({\Box}\hat{P} R)
-\frac{1}{3024} {\Box}^2\hat{P} R
+\frac{1}{3780}\hat{P}{\Box}^2 R
                     \nonumber\\&&\mbox{}
+\frac{1}{3780}{\Box}^2( \hat{P} R)
+\frac{1}{15120}{\Box}\hat{P}{\Box} R
+\frac{1}{3780}{\Box}(\hat{P}{\Box} R)
                        \nonumber\\&&\mbox{}
+\frac{1}{2520}{\Box}(
\nabla_{\mu}\hat{\cal R}^{\mu\nu}
\nabla^{\alpha}\hat{\cal R}_{\alpha\nu})
+\frac{1}{2520}
\nabla_{\mu}\hat{\cal R}^{\mu\nu}\Box
\nabla^{\alpha}\hat{\cal R}_{\alpha\nu}
+\frac{1}{2520}\Box
\nabla_{\mu}\hat{\cal R}^{\mu\nu}
\nabla^{\alpha}\hat{\cal R}_{\alpha\nu}
                    \nonumber\\&&\mbox{}
-\frac{1}{630}
\big[\Box\nabla_{\alpha}\hat{P},\nabla_{\beta}
\hat{\cal R}^{\beta\alpha}\big]
-\frac{1}{630} \Box
\big[\nabla_{\alpha}\hat{P},\nabla_{\beta}
\hat{\cal R}^{\beta\alpha}\big]
-\frac{1}{1260}
\big[\nabla_{\alpha}\hat{P},\Box\nabla_{\beta}
\hat{\cal R}^{\beta\alpha}\big]
\nonumber\\&&\mbox{}
+\frac{1}{840} \Box \nabla_\mu \nabla_\nu \hat{P} R^{\mu\nu}
+\frac{1}{2520} \nabla_\mu \nabla_\nu \hat{P} \Box R^{\mu\nu}
+\frac{1}{840} \Box( \nabla_\mu \nabla_\nu \hat{P} R^{\mu\nu})
                          \nonumber\\&&\mbox{}
+\left[\frac{1}{12600}
{\Box}^2   (R^{\mu\nu\alpha\beta} R_{\mu\nu\alpha\beta})
+\frac{1}{3150}
{\Box}
( R^{\mu\nu\alpha\beta}\nabla_\mu \nabla_\alpha R_{\nu\beta})
+\frac{1}{9450}
R^{\mu\nu\alpha\beta}
\Box
\nabla_\mu \nabla_\alpha R_{\nu\beta}
                       \right.\nonumber\\&&\mbox{}
+\frac{1}{12600}
\nabla_\alpha \nabla_\beta R_{\mu\nu}
\nabla^\mu \nabla^\nu R^{\alpha\beta}
-\frac{1}{6300}\Box
(\nabla_\alpha R_{\mu\nu} \nabla^{\mu}R^{\nu\alpha})
+\frac{1}{18900}
\nabla_\alpha R_{\mu\nu}
\Box
\nabla^{\mu}R^{\nu\alpha}
                             \nonumber\\&&\mbox{}
+\frac{1}{15120}
\Box
(\nabla_\mu \nabla_\nu R R^{\mu\nu})
+\frac{1}{15120}
\Box\nabla_\mu \nabla_\nu R R^{\mu\nu}
+\frac{1}{75600}
\nabla_\mu \nabla_\nu R \Box R^{\mu\nu}
                               \nonumber\\&&\mbox{}
+\frac{1}{50400}
{\Box}^2
(R^{\mu\nu} R_{\mu\nu})
-\frac{11}{75600}
{\Box}
(R^{\mu\nu} \Box R_{\mu\nu})
+\frac{1}{37800}
R^{\mu\nu}{\Box}^2 R_{\mu\nu}
                            \nonumber\\&&\mbox{}
-\frac{1}{15120}
{\Box}
R^{\mu\nu}{\Box} R_{\mu\nu}
 -\frac{1}{56700}  {\Box}^2 R R
+\frac{1}{453600}  {\Box}( {\Box} R R )
                           \nonumber\\&&\left.\mbox{}
-\frac{1}{453600}  {\Box}R  {\Box}R
+\frac{1}{129600}  {\Box}^2(R R)
     \right]\hat{1}
+{\rm O}[\Re^3].  \label{a4}
\end{eqnarray}
We have checked that terms of (\ref{a4}) containing matter
fields are equivalent to a reduced form of this coefficient
found in \cite{Avramidi}.
The expression in the square brackets
disagrees with $a_4$ of  Ref. \cite{AmsterBO89}
in several of its coefficients.
It is straightforward to get from (\ref{a4})
the functional trace $\int d x {\rm tr} \hat{a}_4$
known from \cite{cpt4,JMP94,BGO90}.
Even after implementation of the trace there
remain differences with Ref. \cite{AmsterBO89}
in the quadratic terms.

Needless to say that from
(\ref{K-K}) a similar relationship for the Schwinger-DeWitt
coefficients (\ref{SDs}) follows \cite{Ball89}
\begin{equation}
                          {\hat a}_{n-1}(x,x)=
                g^{-1/2} \frac{\delta}{\delta \hat{P}}
                           \int dx \, g^{1/2}{\rm tr} \,
                            {\hat a}_n(x,x).
\end{equation}


\section{One-loop currents  in  lowest order in the curvature}

Now it would be  a simple task to proceed from heat kernels
(\ref{D..DK}) to universal functional traces (\ref{UFT})
with aid of the proper time equation (\ref{Greenf}).
Of course, according to the generating function principle we
could take as a generating expression the four dimensional
one-loop effective action  (\ref{efac})
which
looks like \cite{cpt4,JMP94}
\begin{eqnarray}
-W&=&  \frac1{2(4\pi)^2}
\int\! dx\, g^{1/2}\, \,{\rm tr}\,
\left\{
\sum^{5}_{i=1} \gamma_i (-\Box_2)\Re_1\Re_2(i)
\right.
\nonumber\\&&\mbox{}
\left.
+\sum^{29}_{i=1}\Gamma_i              \label{W3}
(-\Box_1,-\Box_2,-\Box_3)
\Re_1\Re_2\Re_3(i)
+{\rm  O}[\Re^4]\right\}
\end{eqnarray}
where the second order  and third order
form factors generated from ones of (\ref{TrK3})
by the proper time integration,
and  ultraviolet divergences
within $\gamma_i(-\Box)$  are extracted by
the dimensional regularization at
$\omega\rightarrow 2$,
while
$\Gamma_i(-\Box_1,-\Box_2,-\Box_3)$
are proved to be finite \cite{cpt4,JMP94}.

Either  of these two ways gives the same
result \cite{BGusev92}
\begin{eqnarray}
\frac{\hat{1}} {F(\nabla)} \delta (x,y) |_{y=x} &=&
\frac1{16 \pi^2}\,  g^{1/2}
\left[-\gamma(-\Box)\hat{P} +\frac1{18} R\hat{1}\right] +{\rm O}[\Re^2],
\label{Green1}
\\
\nabla_{\alpha}   \frac{\hat{1}} {F(\nabla)}   \delta(x,y)|_{y=x}
&=&
\frac1 { 16  \pi^2 } \,  g^{1/2}
  \left[
\frac16
\Big( \gamma(-\Box) +
\frac23 \Big) \nabla^{\beta}  \hat{\cal R}_{\beta\alpha}
-  \frac12          \gamma(-\Box)           \nabla_{\alpha}\hat{P}
\right.
\nonumber \\&&
\hphantom{\frac1 { 16  \pi^2 } \,  g^{1/2}}
\left.\mbox{}
+  \frac1{36}        \nabla_{\alpha}R\hat{1}
\right]
+ {\rm O}[\Re^2],
\label{Green2}
\\
\nabla_{\alpha}\nabla_{\beta}\frac{\hat{1}} {F(\nabla)}\delta(x,y)|_{y=x}
&=&
\frac1{16 \pi^2}\,  g^{1/2}
\left[
\frac1{12}
         \Big( \gamma(-\Box) + \frac23 \Big)
          g_{\alpha\beta} \Box  \hat{P}
\right.
\nonumber\\&&\hspace{-20mm}\mbox{}
-\frac1{60}
           \Big( \gamma(-\Box) + \frac{16}{15} \Big)
            \Box R_{\alpha\beta} \hat{1}
+ \frac1{360}
          \Big( \gamma(-\Box) +\frac1{15} \Big)
           g_{\alpha\beta} \Box R \hat{1}
\nonumber \\&&
\hspace{-20mm}\mbox{}
+ \frac16 \Big( \gamma(-\Box) +\frac23 \Big)
        \nabla_{(\alpha} \nabla^{\mu} \hat{\cal R}_{\mu\beta)}
-\frac13 \Big( \gamma(-\Box) + \frac16 \Big)
          \nabla_\alpha \nabla_\beta \hat{P}
\nonumber \\ && \hspace{-20mm}
\left.\mbox{}
+\frac1{180}
          \Big( \gamma(-\Box)  +\frac{61}{15} \Big)
          \nabla_{\alpha} \nabla_\beta R \hat{1}
       \right]  + {\rm O}[\Re^2].
\label{Green3}
\end{eqnarray}
The basic form factor $\gamma(-\Box)$ is derived from
the basic form factor $f(-s\Box)$ (\ref{smallf})
and possesses divergent and finite nonlocal pieces
\[
\gamma(-\Box)=
\frac{1}{2-\omega} +\ln 4 \pi + 2 + {\bf C} -\ln (-\Box)
+{\rm O}[2-\omega], \ \
\omega \rightarrow 2,
\]
${\bf C}$ is  the Euler constant.
The divergences of (\ref{Green1})--(\ref{Green3})
coincide with ones obtained earlier \cite{BV87}.

Let us note that the metric variation of the effective action
itself is of great importance since it is nothing but
the expectation value of the energy-momentum tensor
\cite{cpt,GAVStrasburg}
\[
\Big<
T_{\alpha\beta}
\Big>
\equiv
2 g^{-1/2} \frac{\delta W}{\delta g^{\alpha\beta}}.
\]
Having (\ref{Green3}) in explicit form gives us
$\Big< T_{\alpha\beta}\Big> $
 as well.

The spectral representation $\gamma(-\Box)$
is of the form
\begin{equation}
        \gamma(-\Box)  =  \int_0^\infty
    \Big(\frac{1}{m^2-\Box} - \frac{1}{m^2-\mu^2} \Big),
\end{equation}
where the parameter $\mu^2 > 0$ describes the
ultraviolet renormalization arbitrariness.
These one-loop Euclidean radiation currents
should be analytically continued
to the  physical Lorentzian
spacetime \cite{cpt}.
In the Lorentzian spacetime the unique
Euclidean Green function with zero boundary conditions at
infinity corresponds
to two different setups having the {\it in--out}
 and {\it in--in} ({\it out--out}) vacuum boundary
conditions. The analytic continuation
consists in substituting
\hbox{$1/(m^2-\Box)$ }
by the Feynman or retarded (advanced)
Green functions \cite{cpt,GAVStrasburg}.
The spectral representations for form factors of second
order Green functions (read -- third order effective action)
even in Euclidean spacetime is rather involved and
given in full account in Ref. \cite{cpt4}.

The universal functional traces (\ref{Green1})--(\ref{Green3}) themselves
are generating expressions for polarization operators (\ref{PO})
as seen from
\[
\delta \Big( \frac{1}{F}\Big)=
- \frac{1}{F} \delta F \frac{1}{F}.
\]
The curvature power is again reduced by one
and resulting expressions are
\begin{eqnarray*}
&&
\frac{\delta^{A}_{B}}{F(\nabla)} \, \delta(x,y)\,
 \frac{\delta^{C}_{D}}{F(\nabla)} \,  \delta(x,y)
= \frac{g^{1/2}}{16 \pi^2} \,
    \gamma(-\Box) \, \delta^A_B \, \delta^C_D\, \delta(x,y) \,
          + {\rm O}[\Re],
\\ &&
\nabla_{\alpha}
\frac{\delta^{A}_{B}}{F(\nabla)}\,  \delta(x,y)\,
\frac{\delta^{C}_{D}}{F(\nabla)} \,  \delta(x,y)
= \frac{g^{1/2}}{16 \pi^2} \,
   \frac12 \, \gamma(-\Box) \, \delta^A_B\,  \delta^C_D\,
\nabla_{\alpha}  \delta(x,y)
         + {\rm O}[\Re],
\end{eqnarray*}
and three more are readily computed \cite{BGusev92}
\begin{eqnarray*}
&&\nabla_{\alpha} \nabla_{\beta}
\frac{\delta^{A}_{B}}{F(\nabla)}  \delta(x,y) \,
\frac{\delta^{C}_{D}}{F(\nabla)}   \delta(x,y),
\ \
\nabla_{\alpha}
\frac{\delta^{A}_{B}}{F(\nabla)}  \delta(x,y) \,
\nabla_{\beta}
\frac{\delta^{C}_{D}}{F(\nabla)}   \delta(x,y),
\\
&&
\nabla_{\alpha} \nabla_{\beta}
\frac{\delta^{A}_{B}}{F(\nabla)}  \delta(x,y) \,
\nabla_{\mu}
\frac{\delta^{C}_{D}}{F(\nabla)}   \delta(x,y).
\end{eqnarray*}
Again divergences of the expressions above were
established in \cite{BV87}.
These and other polarization operators
are made up of derivatives and basic form factors
$\gamma(-\Box)$ acting on delta functions.

At this point we can observe an elegant
hierarchy
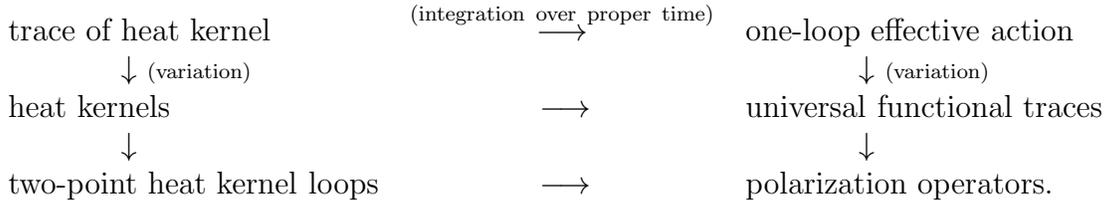
\vspace{5mm}

\begin{tabular}{lll}
trace of heat kernel
& $ \stackrel{\rm (integration\ over\ proper\ time)}{\longrightarrow}$
                         &  one-loop effective action    \\
\hphantom{trace of}
$\downarrow$ {\scriptsize (variation)}
       &
                          &
\hphantom{trace of}
$\downarrow$  {\scriptsize (variation)}   \\
heat kernels
        & \hphantom{derivativ}
 ${\longrightarrow}$
                          &  universal functional traces \\
\hphantom{trace of}
$\downarrow$
         &
                           &
\hphantom{trace of}
$\downarrow$\\
two-point heat kernel loops
          &  \hphantom{derivativ} ${\longrightarrow}$
                            & polarization operators.
\end{tabular}


\section{Conclusions}

In this report we have aimed to outline
the interplay among the whole class
of quantum  objects  including one-loop effective action,
universal functional traces, and polarization operators
whose properties are entirely defined by the
trace of the heat kernel  derived from the covariant
perturbation theory \cite{cpt4}.

We have shown that the generating function method
is simple and efficient while dealing with
elements of two-loop graphs
in quantum gravity and gauge theories.
In this way some ``tadpoles'' and polarization operators have been
obtained as well as heat kernels themselves.
However to start we should be provided by the generating
functions ${\rm Tr} K(s)$ and $W$ with qualitatively
higher accuracy of
perturbation theory. Its present state is limited
to third order in curvature
and any  further progress seems  formidable and unnecessary
in quantum gravity at least, though
in pure gauge field theory  next orders might be manageable
and needed for measurements at hadron colliders such as one-loop
two-quark $n$-gluon QCD amplitudes
\cite{Bern93}.

Another disadvantage of the method  is its inability
to produce Green function with more than two
derivatives since  the operator $F(\nabla)$
contains no more than two of them  acting on
the metric (\ref{oper}).
But the ordinary covariant perturbation theory encompasses
all possible cases of  (\ref{D..DK}).

We have presented second order manipulations only
for the heat kernel (undifferentiated Green function)
 but to reproduce just two-loop local divergences
requires such knowledge for all the
discussed radiation currents as well as for
Green functions with mixed derivatives \cite{BV87,VdV92}.
Despite the fact that our program is far from being complete,
we have indicated here its basic features.

Our concluding remark is that
now the covariant perturbation theory approach
and correspondingly all derivations above
look even more promising in view of the fact  that the same
conventional perturbational rules were obtained for
low-energy limit of string theory \cite{Bern93}.

\section*{Acknowledgments}
We wish to thank Professor T.A. Osborn
for warm hospitality at the University of Manitoba and
support.
A.B. also gratefully acknowledges a Professional Associateships
awarded
by the University of Manitoba Office of Research Administration
as well as support from Russian Foundation for Fundamental Research
under the grant 93-02-15594 and the grant INTAS-93-439 from
European Community Commission.



\begin{thebibliography}{99}
\bibitem{cpt}
                     A.O. Barvinsky and G.A. Vilkovisky,
                        Nucl. Phys. B
                                {\bf 282}, 163  (1987);
                                   {\bf 282}, 471  (1990);
                                     {\bf 333}, 512  (1990).
\bibitem{cpt4}
                       A.O. Barvinsky, Yu.V. Gusev, V.V. Zhytnikov,
and G.A. Vilkovisky,
                         Covariant perturbation theory (IV).
Third order in the curvature,
                         Report of the University of Manitoba
                           (University of Manitoba, Winnipeg, 1993).
\bibitem{JMP94}
                          A.O. Barvinsky, Yu.V. Gusev,  G.A. Vilkovisky,
and V.V. Zhytnikov,
                         J. Math.  Phys.
                            {\bf35}, 3525 (1994), gr-qc/9404061;
                         J. Math. Phys.
                             {\bf 35}, 3543  (1994), gr-qc/9404063;
                         Nucl. Phys. B
                          {\bf 439}, 561 (1995), hep-th/9404187.
\bibitem{GAVStrasburg}
                                    G.A.Vilkovisky,
Heat kernel: rencontre entre physiciens et mathematiciens,
                                   {\em in}
{Publication de l'Institut de
                                 Recherche Math\'ematique Avanc\'ee}, R.C.P.
                                  25, vol.43 (Strasbourg, 1992) p.203,
                                  CERN-TH.6392/92.
\bibitem{DeWittBook}
                                    B.S. DeWitt,
                            {\em Dynamical theory of groups and fields}
                                    (Gordon and Breach, New York, 1965).
\bibitem{BV85}
                             A.O.Barvinsky and G.A.Vilkovisky,
                             Phys. Rep. {\bf  119}, 1 (1985).
\bibitem{BV87}
                         A.O.Barvinsky and G.A.Vilkovisky,
                         {\em in} { Quantum Field Theory and
                             Quantum Statistics},  vol. 1,
                      eds. I.A. Batalin, C.J. Isham and G.A. Vilkovisky
                             (Hilger, Bristol, 1987) p. 245.
\bibitem{VdV92}
                            A.E.M. Van de Ven,
                             Nucl. Phys. B
                            {\bf 378} , 309 (1992).
\bibitem{BGusev92}
                                A.O. Barvinsky and Yu.V. Gusev,
                                Class. Quantum Grav.
                                {\bf 9}, 383 (1992).
\bibitem{Gilkey75}
                             P.B. Gilkey,
                             J.Diff.Geom.
                              {\bf 10}, 601  (1975).
\bibitem{WFO81}
                            S.F.J. Wilk, Y. Fujiwara, and T.A. Osborn,
                             Phys. Rev. A
                              {\bf 24}, 2187 (1981).
\bibitem{Avramidi}
                          I.G. Avramidi,
                         Phys. Lett. B
                                           {\bf 238}, 92 (1990);
                          Nucl. Phys. B
                                             {\bf 355}, 712  (1991).
\bibitem{AmsterBO89}
                                   P. Amsterdamski, A.L. Berkin,
and D.J. O'Connor,
                                    Class. Quantum Grav.
                                    {\bf 6}, 1981  (1989).
\bibitem{BGO90}
                   T.P. Branson, P.B. Gilkey, and B. {\protect \O}rsted,
                   Proc. Am. Math. Soc.
                   {\bf 109}, 437 (1990).
\bibitem{Ball89}
                         R.D. Ball,
                          Phys. Rep.
                          {\bf 182}, 1 (1989).
\bibitem{Bern93}
                           Z. Bern,
                           {\em in} { Recent directions in particle theory},
                           Proceedings of TASI school
                            (Boulder, 1992) p. 471,
                           hep-ph/9304249.
\end{thebibliography}
\end{document}